\documentclass[showpacs,aps,graphicx,twocolumn]{revtex4}
\usepackage{graphicx}

\begin{document}


\title{Nonlocal entanglement concentration scheme for partially entangled
multipartite systems  with  nonlinear optics\footnote{Published in
Phys. Rev. A 77, 062325 (2008)}}

\author{Yu-Bo Sheng,$^{1,2,3}$ Fu-Guo Deng,$^{1,4}$\footnote{Email address:
fgdeng@bnu.edu.cn}  and Hong-Yu Zhou$^{1,2,3}$}
\address{$^1$The Key Laboratory of Beam Technology and Material
Modification of Ministry of Education, Beijing Normal University,
Beijing 100875,  China\\
$^2$Institute of Low Energy Nuclear Physics, and Department of
Material Science and Engineering, Beijing Normal University,
Beijing 100875, China\\
$^3$Beijing Radiation Center, Beijing 100875,  China\\
$^4$Department of Physics, Applied Optics Beijing Area Major
Laboratory, Beijing Normal University, Beijing 100875, China }
\date{\today }

\begin{abstract}

We present a nonlocal entanglement concentration scheme for
reconstructing some maximally entangled multipartite states from
partially entangled ones by exploiting cross-Kerr nonlinearities
to distinguish the parity of  two polarization photons. Compared
with the entanglement concentration schemes based on two-particle
collective unitary evolution, this scheme does not require the
parties to know accurately information about the partially
entangled states---i.e., their coefficients. Moreover, it does not
require the parties to possess  sophisticated single-photon
detectors, which makes this protocol  feasible with present
techniques. By iteration of entanglement concentration processes,
this scheme has a higher efficiency and yield than those with
linear optical elements. All these advantages  make this scheme
more efficient and more convenient than others in  practical
applications.
\end{abstract}
 \maketitle

\section{introduction}

Entanglement is a unique phenomenon in quantum mechanics and it
plays an important role in quantum-information processing and
transmission. For instance, quantum computers exploit entanglement
to speedup the computation of  problems in mathematics
\cite{computation1,computation2}. The two legitimate users in
quantum communication---say, the sender Alice and the receiver
Bob---use an entangled quantum system to transmit a private key
\cite{Ekert91,BBM92,rmp,LongLiu,CORE} or a secret message
\cite{QSDC}. Also quantum dense coding \cite{densecoding,super2},
quantum teleportation \cite{teleportation}, controlled
teleportation \cite{cteleportation}, and quantum-state sharing
\cite{QSTS} need entanglements to set up the quantum channel.
However, in a practical transmission or the process for storing
quantum systems, we can not avoid   channel noise, which will make
the entangled quantum system  less entangled. For example, the
Bell state $\vert \phi^+\rangle_{AB}=\frac{1}{\sqrt{2}}(\vert
H\rangle_A\vert H\rangle_B + \vert V\rangle_A \vert V\rangle_B)$
may become a mixed one such as a Werner state \cite{werner}:
\begin{eqnarray}
W_{F} &=& F \vert \phi^{+}\rangle\langle\phi^{+}\vert + \frac{1-F}{3}(\vert
\phi^{-}\rangle\langle\phi^{-}\vert \nonumber\\
&+& \vert \psi^{+}\rangle\langle\psi^{+} \vert + \vert
\psi^{-}\rangle\langle\psi^{-} \vert),
\end{eqnarray}
where
\begin{eqnarray}
\vert \phi^{\pm} \rangle_{AB} =\frac{1}{\sqrt{2}}(\vert H\rangle_A
\vert H \rangle_B \pm \vert V\rangle_A\vert V\rangle_B),\\
\vert \psi^{\pm} \rangle_{AB} =\frac{1}{\sqrt{2}}(\vert H\rangle_A
\vert V \rangle_B \pm \vert V\rangle_A\vert H\rangle_B).
\end{eqnarray}
Here $H$ and $V$ represent the  horizontal and vertical
polarizations of photons, respectively. The Bell state $\vert
\phi^+\rangle$ can also be degraded as a less pure entangled state
like $|\Psi\rangle=\alpha\vert H\rangle_A \vert V\rangle_B
+\beta\vert V\rangle_A \vert H\rangle_B$, where
$|\alpha|^{2}+|\beta|^{2}=1$. Multipartite entanglement states
also suffer from  channel noise. For instance,
$|\Phi^{\pm}\rangle=\frac{1}{\sqrt{2}}(\vert HH \cdots H\rangle
\pm \vert VV \cdots V\rangle)$ will become
$|\Phi'^{\pm}\rangle=\alpha|HH \cdots H \rangle \pm \beta
|VV\cdots V\rangle$. For three-particle quantum systems, their
states with the form $\vert \Phi^\pm\rangle$ are called
Greenberg-Horne-Zeilinger (GHZ) states. Now, the multipartite
entangled states like
$|\Phi^{\pm}\rangle=\frac{1}{\sqrt{2}}(|00\cdot\cdot\cdot0\rangle\pm|11\cdot\cdot\cdot1\rangle)$
are also called multipartite GHZ states.

The method  of distilling  a mixed state into a maximally
entangled state is named  entanglement purification, which has
been widely studied in recent years \cite{C.H.Bennett1,D.
Deutsch,Pan1,Pan2,M. Murao,M. Horodecki,Yong,shengpra}. Another
way of distilling less entangled pure states into maximally
entangled states that will be detailed here is called entanglement
concentration. Several entanglement concentration protocols of
pure nonmaximally entangled states have been proposed recently.
The first entanglement concentration protocol was proposed by
Bennett \emph{et al} \cite{C.H.Bennett2} in 1996, which is called
Schmidt projection method. In their protocol \cite{C.H.Bennett2},
the two parties of quantum communication need some collective and
nondestructive measurements of photons, which, however, are not
easy to  manipulate in experiment. Also the two parties should
know accurately the coefficients $\alpha$ and $\beta$ of the
partially entangled state $\alpha|01\rangle+\beta|10\rangle$
before entanglement concentration. That is, their protocol works
under the condition that the two users obtain   information about
the coefficients and possess the collective and nondestructive
measurement technique. Another similar scheme is called
entanglement swapping \cite{swapping1,swapping2}. In these schemes
\cite{swapping1,swapping2}, two pairs of less entangled pairs
belong to Alice and Bob. Then Alice sends one of her particles to
Bob, and Bob performs a Bell-state measurement on one of his
particle and Alice's one. So Bob has to own three photons of two
pairs, and they have to perform collective Bell-state
measurements. Moreover, the parties should exploit a two-particle
collective unitary evaluation of the quantum system and an
auxiliary particle to project the partially entangled state into a
maximally entangled one probabilistically.

Recently, two protocols of entanglement concentration  based on a
polarization beam splitter (PBS) were proposed independently by
Yamamoto \emph{et al} \cite{Yamamoto} and Zhao \emph{et al}
\cite{zhao1}. The experimental demonstration of the latter has
been reported \cite{zhao2}. In their protocol, the parties exploit
two PBSs to complete the parity-check measurements of polarization
photons. However, each of the two users Alice and Bob has to
choose the instances in which each of the spatial modes contains
exactly one photon. With current technology,  sophisticated
single-photon detectors are not likely to be available, which
makes it such that these schemes can not be accomplished simply
with linear optical elements.

Cross-Kerr nonlinearity is a powerful tool to construct a
nondestructive quantum nondemolition detector (QND). It also has
the function of constructing a controlled-not (CNOT) gate and a
Bell-state analyzer \cite{QND1}. Cross-Kerr nonlinearity was
widely studied in the generation of qubits
\cite{qubit1,qubit2,qubit3} and the discrimination of unknown
optical qubits \cite{discriminator}. Cross-Kerr nonlinearities can
be described with the Hamiltonian $H_{ck}=\hbar\chi
a^{+}_{s}a_{s}a^{+}_{p}a_{p}$ \cite{QND1,QND2}, where $a^{+}_{s}$
and $a^{+}_{p}$ are the creation operations and   $a_{s}$ and
$a_{p}$ are the destruction operations. If we consider a coherent
beam in the state $|\alpha\rangle_{p}$ with a signal pulse in the
Fock state $|\Psi\rangle_s=c_{0}|0\rangle_{s}+c_{1}|1\rangle_{s}$
($|0\rangle_{s}$ and $|1\rangle_{s}$ denote that there are no
photons and one photon, respectively, in this state), after the
interaction with the cross-Kerr nonlinear medium the whole system
evolves as
\begin{eqnarray}
U_{ck}|\Psi\rangle_{s}|\alpha\rangle_{p}&=& e^{iH_{ck}t/\hbar}[c_{0}|0\rangle_{s}+c_{1}
|1\rangle_{s}]|\alpha\rangle_{p} \nonumber\\
                                        &=& c_{0}|0\rangle_{s}|\alpha\rangle_{p}+c_{1}|1\rangle_{s}|
                                        \alpha
                                        e^{i\theta}\rangle_{p},
\end{eqnarray}
where $\theta=\chi t$ and $t$ is the interaction time. From this
equation, the coherent beam picks up a phase shift directly
proportional to the number of  photons in the Fock state
$|\Psi\rangle_s$. This good feature can be used to construct a
parity-check measurement device \cite{QND1}.

In this paper, we present a different scheme for nonlocal
entanglement concentration of partially entangled multipartite
states with cross-Kerr nonlinearities. By exploiting a new
nondestructive QND, the parties of quantum communication can
accomplish entanglement concentration efficiently without
sophisticated single-photon detectors. Compared with the
entanglement concentration schemes based on linear optical
elements  \cite{Yamamoto,zhao1}, the present scheme has a higher
efficiency and yield. Moreover, it does not require that the
parties know accurately information about the partially entangled
states---i.e., the coefficients of the states---different from
schemes based on two-particle collective unitary evaluation
\cite{C.H.Bennett2,swapping1,swapping2}. These good features give
this scheme  the advantage of high efficiency and  feasibility in
practical applications.

\begin{figure}[!h]
\begin{center}
\includegraphics[width=6cm,angle=0]{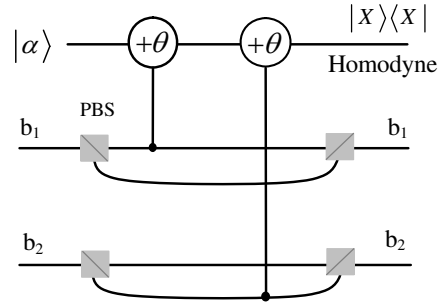}
\caption{The principle of our nondestructive quantum nondemolition
detector (QND). Two cress-Kerr nonlinearities are used to
distinguish superpositions and mixtures of  $|HH\rangle$ and
$|VV\rangle$ from $|HV\rangle$ and $|VH\rangle$. Each polarization
beam splitter (PBS) is used to pass through $|H\rangle$
polarization photons and reflect  $|V\rangle$ polarization
photons. Cross-Kerr nonlinearity will cause the coherent beam to
pick up a phase shift $\theta$ if there is a photon in the mode.
So the probe beam $\vert \alpha \rangle$ will pick up a phase
shift of $\theta$ if the state is $|HH\rangle$ or $|VV\rangle$.
Here $b_1$ and $b_2$ represent the up spatial mode and the down
spatial mode, respectively.}
\end{center}
\end{figure}

\section{entanglement concentration of pure entangled photon pairs}

\subsection{Primary entanglement concentration of less entangled photon pairs }
\label{ecp}

The principle of our nondestructive QND is shown in Fig.1. It is
made up of four PBSs, two identical cross-Kerr nonlinear media,
and an $X$ homodyne measurement. If two polarization photons are
initially prepared in the states $|\varphi\rangle_{b_1}=c_{0}|H
\rangle_{b_1}+c_{1}|V\rangle_{b_1}$ and
$|\varphi\rangle_{b_2}=d_{0}|H\rangle_{b_2}+d_{1}|V\rangle_{b_2}$,
the two photons combined with a coherent beam whose initial state
is $\vert \alpha \rangle_p$ interact with cross-Kerr
nonlinearities, which will evolve the state of the composite
quantum system from the original one $\vert \Psi\rangle_{O}=\vert
\varphi\rangle_{b_1}\otimes \vert \varphi\rangle_{b_2}\otimes
\vert \alpha\rangle_{p}$ to
\begin{eqnarray}
|\Psi\rangle_{T} &=& [c_{0}d_{0}|HH\rangle +
c_{1}d_{1}|VV\rangle]|\alpha e^{i\theta}\rangle_{p} \nonumber\\
&+& c_{0}d_{1}|HV\rangle|\alpha
e^{i2\theta}\rangle_{p}+c_{1}d_{0}|VH\rangle|\alpha\rangle_{p}.
\end{eqnarray}
One can find immediately that   $|HH\rangle$ and $|VV\rangle$
cause the coherent beam $\vert\alpha\rangle_p$ to pick up a phase
shift $\theta$, $|HV\rangle$ to pick up a phase shift $2\theta$,
and $|VH\rangle$ to pick up no phase shift. The different phase
shifts can be distinguished by a general homodyne-heterodyne
measurement ($X$ homodyne measurement). In this way,  one can
distinguish $|HH\rangle$ and $|VV\rangle$ from $|HV\rangle$ and
$|VH\rangle$. This device is also called a two-qubit polarization
parity QND detector. Our QND shown in Fig.1 is a little different
from the one proposed by Nemoto and Munro \cite{QND1}. With the
QND in \cite{QND1}, the $|HH\rangle$ and $|VV\rangle$ pick up no
phase shift. However, it is well known that a vacuum state
(zero-photon state) can also cause there to be no phase shift on
the coherent beam. So one can not distinguish whether two photons
or no photons pass through the two spatial modes. This modified
QND can exactly check the number of photons if they have the same
parity.

\begin{figure}[!h]
\begin{center}
\includegraphics[width=8cm,angle=0]{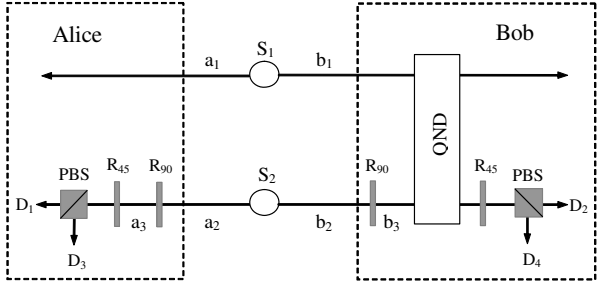}
\caption{Schematic diagram of the proposed entanglement
concentration protocol. Two pairs of identical less entanglement
photons are sent to Alice and Bob from source 1 ($S_1$) and source
2 ($S_2$). The QND is a parity-checking device. The wave plates
$R_{45}$ and $R_{90}$ rotate the horizontal and vertical
polarizations by $45^{\circ}$ and $90^{\circ}$ respectively.}
\end{center}
\end{figure}

With the QND shown in Fig.1, the principle of our entanglement
concentration protocol is shown in Fig.2. Suppose there are two
identical photon pairs with less entanglement $a_1b_1$ and
$a_2b_2$. The photons $a$ belong to Alice and  photons $b$ to Bob.
The photon pairs $a_1b_1$ and $a_2b_2$ are initially in the
following unknown polarization entangled states:
\begin{eqnarray}
|\Phi\rangle_{a_1b_1}=\alpha|H\rangle_{a_1}|H\rangle_{b_1}+\beta|V\rangle_{a_1}|V\rangle_{b_1},\nonumber\\
|\Phi\rangle_{a_2b_2}=\alpha|H\rangle_{a_2}|H\rangle_{b_2}+\beta|V\rangle_{a_2}|V\rangle_{b_2},
\end{eqnarray}
where $|\alpha|^{2}+|\beta|^{2}=1$. The original state of the four
photons can be written as
\begin{eqnarray}
|\Psi\rangle &\equiv& |\Phi\rangle_{a_1b_1}\otimes
|\Phi\rangle_{a_2b_2}=\alpha^{2}|H\rangle_{a_1}|H\rangle_{b_1}|H\rangle_{a_2}|H\rangle_{b_2}\nonumber\\
&+& \alpha\beta|H\rangle_{a_1}|H\rangle_{b_1}|V\rangle_{a_2}|V\rangle_{b_2}\nonumber\\
&+&\alpha\beta|V\rangle_{a_1}|V\rangle_{b_1}|H\rangle_{a_2}|H\rangle_{b_2}\nonumber\\
&+&
\beta^{2}|V\rangle_{a_1}|V\rangle_{b_2}|V\rangle_{a_2}|V\rangle_{b_2}.
\end{eqnarray}
After the two parties Alice and Bob rotate the polarization states
of their second photons $a_2$ and $b_2$ by $90^{\circ}$ with
half-wave plates (i.e., $R_{90}$ shown in Fig.2), the state of the
four photons can be written as
\begin{eqnarray}
|\Psi\rangle^{'} &=&\alpha^{2}|H\rangle_{a_1}|V\rangle_{a_3}|H\rangle_{b_1}|V\rangle_{b_3}\nonumber\\
&+& \alpha\beta|H\rangle_{a_1}|H\rangle_{a_3}|H\rangle_{b_1}|H\rangle_{b_3}\nonumber\\
&+&
\alpha\beta|V\rangle_{a_1}|V\rangle_{a_3}|V\rangle_{b_1}|V\rangle_{b_3}\nonumber\\
&+&
\beta^{2}|V\rangle_{a_1}|H\rangle_{a_3}|V\rangle_{b_1}|H\rangle_{b_3}.\label{staterotation}
\end{eqnarray}
Here $a_3$ ($b_3$) is used to label the photon $a_2$ ($b_2$) after
the half-wave plate $R_{90}$.

From Eq.(\ref{staterotation}), one can see that the terms
$|H\rangle_{a_1}|H\rangle_{a_3}|H\rangle_{b_1}|H\rangle_{b_3}$ and
$|V\rangle_{a_1}|V\rangle_{a_3}|V\rangle_{b_1}|V\rangle_{b_3}$
have the same coefficient of $\alpha\beta$, but the other two
terms are different. Now Bob lets the two photons $b_1$ and $b_3$
enter into the QND. With his homodyne measurement, Bob may get one
of three different results: $|HH\rangle$ and $|VV\rangle$ lead to
a phase shift of $\theta$ on the coherent beam, $|HV\rangle$ leads
to $2\theta$, and the other is $|VH\rangle$, which leads to no
phase shift. If the phase shift of homodyne measurement is
$\theta$, Bob asks Alice to keep these two pairs; otherwise, both
pairs are removed. After only this parity-check measurement, the
state of the photons remaining  becomes
\begin{eqnarray}
|\Psi\rangle^{''} &=& \frac{1}{\sqrt{2}}(|H\rangle_{a_1}|H\rangle_{a_3}|H\rangle_{b_1}|H\rangle_{b_3}\nonumber\\
&+& |V\rangle_{a_1}|V\rangle_{a_3}|V\rangle_{b_1}|V\rangle_{b_3}).
\label{maxstate}
\end{eqnarray}
The probability that Alice and Bob get the above state  is
$P_{s_1}=2|\alpha\beta|^{2}$.

Now both  pairs $a_1b_1$ and $a_3b_3$ are in the same
polarizations. Alice and Bob use their $\lambda/4$-wave plates
$R_{45}$ to rotate the photons  $a_3$ and $b_3$ by $45^{\circ}$.
The unitary transformation of $45^{\circ}$ rotations can be
described as
\begin{eqnarray}
|H\rangle_{a_3} & \rightarrow & \frac{1}{\sqrt{2}}(|H\rangle_{a_3}+|V\rangle_{a_3}),\nonumber\\
|H\rangle_{b_3} & \rightarrow & \frac{1}{\sqrt{2}}(|H\rangle_{b_3}+|V\rangle_{b_3}),\nonumber\\
|V\rangle_{a_3} & \rightarrow & \frac{1}{\sqrt{2}}(|H\rangle_{a_3}-|V\rangle_{a_3}),\nonumber\\
|V\rangle_{b_3} & \rightarrow &
\frac{1}{\sqrt{2}}(|H\rangle_{b_3}-|V\rangle_{b_3}).
\end{eqnarray}
After the rotations, Eq. (\ref{maxstate}) will evolve into
\begin{eqnarray}
 |\Psi\rangle^{'''}&=&\frac{1}{2\sqrt{2}}(|H\rangle_{a1}|H\rangle_{b1}+|V\rangle_{a1}|V\rangle_{b1})
 (|H\rangle_{a3}|H\rangle_{b3}\nonumber\\
&+&
|V\rangle_{a3}|V\rangle_{b3})+\frac{1}{2\sqrt{2}}(|H\rangle_{a1}|H\rangle_{b1}-
|V\rangle_{a1}|V\rangle_{b1})\nonumber\\
&&
(|H\rangle_{a3}|V\rangle_{b3}+|V\rangle_{a3}|H\rangle_{b3}).\label{statedistinguish}
\end{eqnarray}
The last step is to distinguish the photons $a_3$ and $b_3$ in
different polarizations. Two PBSs are used to pass through
$|H\rangle$ polarization photons and reflect   $|V\rangle$
photons. From the Eq. (\ref{statedistinguish}), one can see that
if the two detectors $D_1$ and $D_2$ or the two detectors $D_3$
and $D_4$ fire, the photon pair $a_1b_1$ is left in the state
\begin{eqnarray}
|\phi^{+}\rangle_{a_1b_1}=\frac{1}{\sqrt{2}}(|H\rangle_{a_1}|H\rangle_{b_1}+|V\rangle_{a_1}|V\rangle_{b_1}).
\end{eqnarray}
If $D_1$ and $D_3$ or $D_2$ and $D_4$ fire, the photon pair
$a_1b_1$ are left in the state
\begin{eqnarray}
|\phi^{-}\rangle_{a_1b_1}=\frac{1}{\sqrt{2}}(|H\rangle_{a_1}|H\rangle_{b_1}-|V\rangle_{a_1}|V\rangle_{b_1}).
\end{eqnarray}
 Both of these two states are the maximally entangled ones. In order to get
 the same state of $|\Phi\rangle^{+}_{a1b1}$, one of the two parties Alice and Bob
 should perform a simple local operation of phase rotation on her or his photon. The
 maximally entangled states are generated with above operations.

In our scheme, only one QND is used to detect the parity of the
two polarization photons. If the two photons are in the same
polarization $|HH\rangle$ or $|VV\rangle$, the phase shift of the
coherent beam is $\theta$, which is easy to detect by the homodyne
measurement. Furthermore, our scheme is not required to have
sophisticated single-photon detectors, but only conventional
photon detectors. This is a good feature of our scheme, compared
with  other schemes.

\subsection{Reusing resource-based entanglement concentration of partially entangled photon pairs}
\label{iterationsection}

With only one QND, our entanglement concentration has the same
efficiency as that based on linear optics \cite{Yamamoto,zhao1}.
The yield of maximally entangled states $Y$ is $|\alpha \beta|^2$.
Here the yield is defined as the ratio of the number of maximally
entangled photon pairs, $N_m$, and the number of  originally less
entangled photon pairs, $N_l$. That is, the yield of our scheme
discussed above is $Y_1=\frac{N_m}{N_l}=|\alpha \beta|^2$. In
fact, $Y_1$ is not the maximal value of the yield of the
entanglement concentration scheme with the QND.

In our entanglement concentration scheme above, the two parties
Alice and Bob only pick up  instances in which Bob gets the phase
shift $\theta$ on his coherent beam and removes the other
instances. In this way, the photon pairs kept are in the state
$\vert \Psi\rangle''$. However, if Bob chooses a suitable
cross-Kerr medium and controls accurately the interaction time
$t$, he can make the phase shift $\theta=\chi t=\pi$. In this way,
$2\theta$ and $0$ represent the same phase shift $0$. The two
photon pairs removed by Alice and Bob in the scheme above are just
in the state
\begin{eqnarray}
|\Phi_1\rangle^{''} &=& \alpha^2
|H\rangle_{a_1}|V\rangle_{a_3}|H\rangle_{b_1}|V\rangle_{b_3}\nonumber\\
&+& \beta^2
|V\rangle_{a_1}|H\rangle_{a_3}|V\rangle_{b_1}|H\rangle_{b_3}.
\label{lessstate2}
\end{eqnarray}
This four-photon system is not in a maximally entangled state, but
it can be used to get some maximally entangled state with
entanglement concentration. In detail, Alice and Bob take a
rotation by $90^{\circ}$ on each photon of the second four-photon
system and cause the state of this system to become
\begin{eqnarray}
|\Phi_2\rangle^{''}&=& \beta^2
|H\rangle_{a'_1}|V\rangle_{a'_3}|H\rangle_{b'_1}|V\rangle_{b'_3}\nonumber\\
&+& \alpha^2
|V\rangle_{a'_1}|H\rangle_{a'_3}|V\rangle_{b'_1}|H\rangle_{b'_3}.
\label{lessstate3}
\end{eqnarray}
The state of the composite system composed of  eight photons
becomes
\begin{eqnarray}
|\Phi_s\rangle^{''} &\equiv& |\Phi_1\rangle^{''}\otimes
|\Phi_2\rangle^{''}\nonumber\\
&=&
\alpha^2\beta^2(|H\rangle_{a_1}|V\rangle_{a_3}|H\rangle_{b_1}|V\rangle_{b_3}
|H\rangle_{a'_1}|V\rangle_{a'_3}|H\rangle_{b'_1}|V\rangle_{b'_3}\nonumber\\
&+&
|V\rangle_{a_1}|H\rangle_{a_3}|V\rangle_{b_1}|H\rangle_{b_3}
|V\rangle_{a'_1}|H\rangle_{a'_3}|V\rangle_{b'_1}|H\rangle_{b'_3})\nonumber\\
&+& \alpha^4
|H\rangle_{a_1}|V\rangle_{a_3}|H\rangle_{b_1}|V\rangle_{b_3}
|V\rangle_{a'_1}|H\rangle_{a'_3}|V\rangle_{b'_1}|H\rangle_{b'_3}\nonumber\\
&+& \beta^4
|V\rangle_{a_1}|H\rangle_{a_3}|V\rangle_{b_1}|H\rangle_{b_3}
|H\rangle_{a'_1}|V\rangle_{a'_3}|H\rangle_{b'_1}|V\rangle_{b'_3}.\nonumber\\
\label{lessstate4}
\end{eqnarray}
For picking up the first two terms, Bob need only detect the
parities of the two photons $b_3$ and $b'_3$ with the QND. As the
two polarization photons $b_3$ and $b'_3$ in the first two terms
have the same parity, they will cause the coherent beam $\vert
\alpha \rangle_p$ to have a phase shift $\theta=\pi$. Those in the
other two terms cause the coherent beam $\vert \alpha \rangle_p$
to have a phase shift $0$.

When Bob gets the phase shift  $\theta=\pi$, the eight photons
collapse to the state
\begin{eqnarray}
|\Phi_s\rangle^{'''} &=&\frac{1}{\sqrt{2}}
(|H\rangle_{a_1}|V\rangle_{a_3}|H\rangle_{b_1}|V\rangle_{b_3}
|H\rangle_{a'_1}|V\rangle_{a'_3}|H\rangle_{b'_1}|V\rangle_{b'_3}\nonumber\\
&+& |V\rangle_{a_1}|H\rangle_{a_3}|V\rangle_{b_1}|H\rangle_{b_3}
|V\rangle_{a'_1}|H\rangle_{a'_3}|V\rangle_{b'_1}|H\rangle_{b'_3}).\nonumber\\
\label{lessstate5}
\end{eqnarray}
The probability that Alice and Bob get this state is
\begin{eqnarray}
P_{s_2}=\frac{2|\alpha\beta|^{4}}{(|\alpha|^4 + |\beta|^4)^2}.
\end{eqnarray}
They have the probability $P'_{f_2}=1  - P_{s_2}$ to obtain the
less entangled state
\begin{eqnarray}
|\Phi_1\rangle^{'''} &=& \alpha^4
|H\rangle_{a_1}|V\rangle_{a_3}|H\rangle_{b_1}|V\rangle_{b_3}
|V\rangle_{a'_1}|H\rangle_{a'_3}|V\rangle_{b'_1}|H\rangle_{b'_3}\nonumber\\
&+& \beta^4
|V\rangle_{a_1}|H\rangle_{a_3}|V\rangle_{b_1}|H\rangle_{b_3}
|H\rangle_{a'_1}|V\rangle_{a'_3}|H\rangle_{b'_1}|V\rangle_{b'_3}\nonumber\\
\label{lessstate6}
\end{eqnarray}
which can be used to concentrate entanglement by iteration of the
process discussed above. In this way, one can obtain easily the
probability
\begin{eqnarray}
P_{s_n}=\frac{2|\alpha\beta|^{2^n}}{(|\alpha|^{2^n} +
|\beta|^{2^n})^2},
\end{eqnarray}
where $n$ is the iteration number of the entanglement
concentration processes.

For the four photons in the state described by
Eq.(\ref{lessstate5}), Alice and Bob can obtain a maximally
entangled photon pair with some single-photon measurements on the
other six photons by choosing the basis $X=\{\vert \pm
x\rangle=\frac{1}{\sqrt{2}}(\vert H\rangle \pm \vert V\rangle)\}$.
That is, Alice and Bob first rotate their polarization photons
$a_3$, $b_3$, $a'_1$, $b'_1$, $a'_3$ and $b'_3$ by $45^{\circ}$,
similar to the case discussed above (shown in Fig.2), and then
measure these six photons. If the number of the antiparallel
outcomes obtained by Alice and Bob is even, the photon pair
$a_1b_1$ collapses to the state $\vert
\phi^+\rangle_{a_1b_1}=\frac{1}{\sqrt{2}}(\vert
H\rangle_{a_1}\vert H\rangle_{b_1} + \vert V\rangle_{a_1}\vert
V\rangle_{b_1})$; otherwise the photon pair $a_1b_1$ collapses to
the state $\vert \phi^-\rangle_{a_1b_1}=\frac{1}{\sqrt{2}}(\vert
H\rangle_{a_1}\vert H\rangle_{b_1} - \vert V\rangle_{a_1}\vert
V\rangle_{b_1})$.

With the iteration of the entanglement concentration process, the
yield of our scheme is improved to be $Y$---i.e.,
\begin{eqnarray}
Y &=& \sum_{i=1}^{n} Y_i,
\end{eqnarray}
where
\begin{eqnarray}
Y_1 &=& |\alpha\beta|^{2},\nonumber\\
Y_2 &=&
\frac{1}{2}(1-2|\alpha\beta|^2)\frac{|\alpha\beta|^4}{(|\alpha|^4
+
|\beta|^4)^2}\nonumber,\\
Y_3&=&
\frac{1}{2^2}(1-2|\alpha\beta|^2)[1-\frac{|\alpha\beta|^4}{(|\alpha|^4
+ |\beta|^4)^2}]\frac{|\alpha\beta|^8}{(|\alpha|^8 +
|\beta|^8)^2},\nonumber\\
&& \;\;\;\;\;\;\;\;\;\;\;\;\;\;\;\;   \ldots \nonumber\\
Y_n &=& \frac{1}{2^{n-1}}
(1-2|\alpha\beta|^2)\left(\prod^{n-1}_{j=3}[1-\frac{2|\alpha\beta|^{2^{j-1}}}{(|\alpha|^{2^{j-1}}
+
|\beta|^{2^{j-1}})^2}]\right)\nonumber\\
&&\frac{|\alpha\beta|^{2^{n}}}{(|\alpha|^{2^n} +
|\beta|^{2^n})^2}.
\end{eqnarray}
The yield is shown in Fig.3 with the change of the iteration number
of entanglement concentration processes $n$  and the coefficient
$\alpha \in[0,\frac{1}{\sqrt{2}}]$.

\begin{figure}[!h]
\begin{center}
\includegraphics[width=8cm,angle=0]{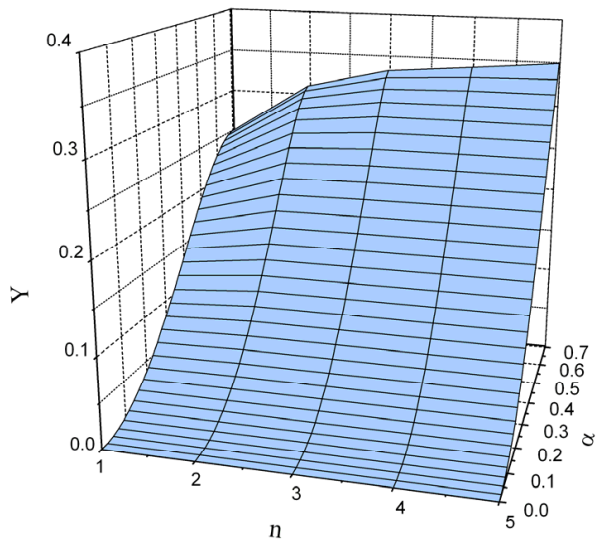}
\caption{(Color online) The yield ($Y$) is altered with the
iteration number of entanglement concentration processes $n$ and the
coefficient $\alpha\in [0,\frac{1}{\sqrt{2}}]$. }
\end{center}
\end{figure}

Certainly, Alice and Bob can also accomplish the iteration of the
entanglement concentration by first measuring the two photons
$a_3$ and $b_3$ in the state $|\Phi_1\rangle^{''}$ described by
 Eq. (\ref{lessstate2}) with the basis $X$ and then
concentrating some maximally entangled states from the partially
entangled quantum systems composed of the pairs $a_1b_1$. In fact,
after the measurements of the two photons with the basis $X$,
Alice and Bob can transfer the state of photon pair $a_1b_1$ to
$\alpha^2 \vert H\rangle_{a_1}\vert H\rangle_{b_1} + \beta^2 \vert
V\rangle_{a_1}\vert V\rangle_{b_1}$ with or without a unitary
operation. Alice and Bob can accomplish the entanglement
concentration with the same way discussed in Sec. \ref{ecp}.

The same as the entanglement concentration schemes with linear
optical elements \cite{Yamamoto,zhao1}, the present scheme has the
advantage that the two parties of quantum communication are not
required to know the coefficients of the less entangled states in
advance in order to reconstruct some maximally entangled states.
Moreover, this scheme does not require sophisticated single-photon
detectors and has a higher yield of maximally entangled states
than those based on linear optical elements \cite{Yamamoto,zhao1}
as the efficiency in the latter is just $|\alpha\beta|^2$ [the
probability that Alice and Bob get an Einstein- Podolsky-Rosen
(EPR) pair   from two partially entangled photon pairs is
$2|\alpha\beta|^2$ in Refs. \cite{Yamamoto,zhao1}]. These good
features make the present entanglement concentration scheme more
efficient and more convenient than others in practical
applications.

\section{entanglement concentration of less entangled multipartite GHZ-class states}

It is straightforward to generalize our entanglement concentration
scheme to reconstruct maximally entangled multipartite GHZ states
from partially entangled GHZ-class states.

Suppose the partially entangled $N$-particle GHZ-class states are
described as follows:
\begin{eqnarray}
|\Phi'^{+}\rangle=\alpha|HH\cdot\cdot\cdot
H\rangle+\beta|VV\cdot\cdot\cdot V\rangle,
\end{eqnarray}
where $|\alpha|^{2}+|\beta|^{2}=1$. For two GHZ-class states, the
composite state can be written as
\begin{eqnarray}
   |\Psi'\rangle &=& |\Phi'^+\rangle_{1}\otimes|\Phi'^+\rangle_{2}=(\alpha|H\rangle_{1}|H\rangle_{2}\cdots
   |H\rangle_{N}\nonumber\\
&&+\beta|V\rangle_{1}|V\rangle_{2}\cdot\cdot\cdot|V\rangle_{N})\otimes\nonumber\\
&&(\alpha|H\rangle_{N+1}|H\rangle_{N+2}\cdot\cdot\cdot|H\rangle_{2N}\nonumber\\
&&+\beta|V\rangle_{N+1}|V\rangle_{N+2}\cdot\cdot\cdot|V\rangle_{2N}).
\end{eqnarray}

\begin{figure}[!h]
\begin{center}
\includegraphics[width=9cm,angle=0]{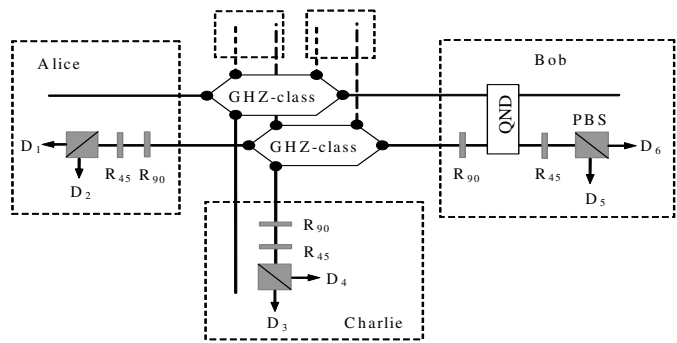}
\caption{Schematic diagram of the multipartite entanglement
concentration scheme. $2N$ particles in two partially entangled
$N$-particle GHZ-class states are sent to $N$ parties of quantum
communication---say Alice, Bob, Charlie, etc. Photons $2$ and
$N+2$ are sent to Bob and enter into QND to  complete a
parity-check measurement. After the QND measurement, Bob asks the
others to retain their photons if his two photons have the same
parity ($|HH\rangle$ or $|VV\rangle$) and remove them for next
iteration if Bob gets an odd parity ($|HV\rangle$ or
$|VH\rangle$). }
\end{center}
\end{figure}

The principle of our entanglement concentration scheme for
multipartite GHZ-class states is shown in Fig.4. $2N$ photons in
two pairs of $N$-particle non-maximally entangled GHZ-class states
are sent to Alice, Bob, Charlie, ect. (i.e., the $N$ parties of
quantum communication). Each party gets two photons. One comes
from the state $|\Phi^+\rangle_{1}$ and the other comes from
$|\Phi^+\rangle_{2}$, shown in Fig.4. Suppose Alice gets photon 1
and the photon $N+1$  and Bob gets   photon $2$ and  photon $N+2$.
Before entanglement concentration, each party rotates his second
polarization photon by $90^{\circ}$, similar to the case for
concentrating two-photon pairs. After the $90^{\circ}$ rotations,
the state of the $2N$ photons becomes
\begin{eqnarray}
 |\Psi'\rangle' &=& \alpha^{2}|H\rangle_{1}|H\rangle_{2}\cdot\cdot\cdot|H\rangle_{N}|V\rangle_{N+1}
 |V\rangle_{N+2}\cdot\cdot\cdot|V\rangle_{2N}\nonumber\\
&+&\alpha\beta|H\rangle_{1}|H\rangle_{2}\cdot\cdot\cdot|H\rangle_{N}|H\rangle_{N+1}|H\rangle_{N+2}\cdots
|H\rangle_{2N}\nonumber\\
&+&\alpha\beta|V\rangle_{1}|V\rangle_{2}\cdot\cdot\cdot|V\rangle_{N}|V\rangle_{N+1}|V\rangle_{N+2}\cdots
|V\rangle_{2N}\nonumber\\
&+&\beta^{2}|V\rangle_{1}|V\rangle_{2}\cdot\cdot\cdot|V\rangle_{N}|H\rangle_{N+1}|H\rangle_{N+2}\cdots
|H\rangle_{2N}.\nonumber\\
\end{eqnarray}
Bob lets  photons $2$ and $N+2$ pass through his QND detector
whose principle is shown in Fig.2. For  $|HH\rangle$ and
$|VV\rangle$, Bob gets the result with an $X$ homodyne measurement
$\theta$; for $|HV\rangle$, the result is $2\theta$ and
$|VH\rangle$ will make no phase shift. By choosing the phase shift
$\theta$, Bob asks the others to retain their photons; otherwise,
all the parties remove the photons. In this way, the whole state
of the retained photons can be described as
\begin{eqnarray}
 |\Psi'\rangle'' &=& \frac{1}{\sqrt{2}}(|H\rangle_{1}|H\rangle_{2}\cdots|H\rangle_{N}
 |H\rangle_{N+1}|H\rangle_{N+2}\cdots|H\rangle_{2N}\nonumber\\
&&+|V\rangle_{1}|V\rangle_{2}\cdots|V\rangle_{N}|V\rangle_{N+1}|V\rangle_{N+2}\cdots
|V\rangle_{2N}).\nonumber\\
\end{eqnarray}
The success probability is $2|\alpha\beta|^2$, the same as that
for two-photon pairs $P_{s_1}$. The above state is a maximally
entangled $2N$-particle state. By measuring each of the photons
coming from the second GHZ-class state with  basis $X$, the
parties will obtain a maximally entangled $N$-particle state, as
after the photons $N+1$, $N+2$, $\ldots$, and $2N$ pass through
the $R_{45}$ plates, which rotate the polarizations of photons by
$45^{\circ}$, the state of the composite system becomes
\begin{eqnarray}
|\Psi'\rangle''' &=&
\frac{1}{\sqrt{2}}[|H\rangle_{1}|H\rangle_{2}\cdot\cdot\cdot|H\rangle_{N}
(\frac{1}{\sqrt{2}})^{\otimes^{N}}(|H\rangle+|V\rangle)^{\otimes^{N}}\nonumber\\
&+&|V\rangle_{1}|V\rangle_{2}\cdot\cdot\cdot|V\rangle_{N}
(\frac{1}{\sqrt{2}})^{\otimes^{N}}(|H\rangle-|V\rangle)^{\otimes^{N}}].\nonumber\\
\end{eqnarray}
By measuring the $N$ photon with the conventional photon
detectors, the $N$ parties will obtain a maximally entangled state
$\vert GHZ^+\rangle_{12\cdots N}$ if the number of parties who
obtain a single-photon measurement outcome $\vert V\rangle$ is
even; otherwise, they will obtain the maximally entangled state
$\vert GHZ^-\rangle_{12\cdots N}$. Here
\begin{eqnarray}
|GHZ^+\rangle &=&
\frac{1}{\sqrt{2}}(|H\rangle_{1}|H\rangle_{2}\cdots|H\rangle_{N}+
|V\rangle_{1}
|V\rangle_{2}\cdots|V\rangle_{N}),\nonumber\\
\end{eqnarray}
and
\begin{eqnarray}
|GHZ^-\rangle &=&
\frac{1}{\sqrt{2}}(|H\rangle_{1}|H\rangle_{2}\cdots|H\rangle_{N}-
|V\rangle_{1}
|V\rangle_{2}\cdots|V\rangle_{N}).\nonumber\\
\end{eqnarray}

For the photons removed by the parties, the method discussed in
Sec. \ref{iterationsection} also works for improving the
efficiency of a successful concentration of GHZ-class states and
the yield. In this time, one need only replace $\vert HH\rangle$
and $\vert VV\rangle$ in Sec. \ref{iterationsection}  with $\vert
HH\cdots H\rangle$ and $\vert VV\cdots V\rangle$, respectively.

\section{discussion and summary}

Compared with the entanglement concentration schemes
\cite{C.H.Bennett2,swapping1,swapping2}  by evolving the composite
system and an auxiliary particle, the present scheme does not
require the parties of quantum communication to know accurately
 information about the less entanglement states. This good
feature makes the present scheme more efficient than those in
Refs. \cite{C.H.Bennett2,swapping1,swapping2} as the decoherence
of entangled quantum systems depends on the noise of quantum
channels or the interaction with the environment, which causes the
two parties to be blind to the information about the state. With
sophisticated single-photon detectors,  entanglement concentration
schemes \cite{Yamamoto,zhao1} with linear optical elements are
efficient for concentrating some partially entangled states. With
the development of technology, sophisticated single-photon
detectors may be obtained in the future even though they are far
beyond what is experimentally feasible at present. Cross-Kerr
nonlinearity provides a good  QND with which a parity-check
measurement can be accomplished perfectly in principle
\cite{QND1}. With the QND, our entanglement concentration scheme
has a higher efficiency and yield than those with linear optical
elements \cite{Yamamoto,zhao1}.

In summary, we propose a different scheme for nonlocal
entanglement concentration of partially entangled multipartite
states. We exploit cross-Kerr nonlinearities to distinguish the
parity of   two polarization photons. Compared with other
entanglement concentration schemes, this scheme does not require a
collective measurement  and does not require the parties of
quantum communication to know the coefficients  $\alpha$ and
$\beta$ of the less entangled states. This advantage makes our
scheme have the capability of distilling arbitrary multipartite
GHZ-class states. Moreover, it does not require the parties to
adopt sophisticated single-photon detectors, which makes this
scheme feasible with present techniques. By iteration of
entanglement concentration processes, this scheme has a higher
efficiency than those with linear optical elements. All these
advantages  make this scheme more convenient in  practical
applications than others.

\section*{ACKNOWLEDGEMENTS}

This work is supported by the National Natural Science Foundation of
China under Grant No. 10604008, A Foundation for the Author of
National Excellent Doctoral Dissertation of China under Grant No.
200723, and Beijing Education Committee under Grant No. XK100270454.

\end{document}